# First experimental study of multiple orientation muon tomography, with image optimization in sparse data environments

Jesus J. Valencia[1], Adam A. Hecht[1], C. L. Morris[2], E. Guardincerri[2], D. Poulson[2], J. Bacon[2], J. M. Durham[2]
1. Department of Nuclear Engineering, University of New Mexico, Albuquerque, NM 87131, USA
2. Los Alamos National Laboratory, Los Alamos, NM 87545, USA

**ABSTRACT**

Due to the high penetrating power of cosmic ray muons, they can be used to probe very thick and dense objects. As charged particles, they can be tracked by ionization detectors, determining the position and direction of the muons. With detectors on either side of an object, particle direction changes can be used to extract scattering information within an object. This can be used to produce a scattering intensity image within the object related to density and atomic number. Such imaging is typically performed with a single detector-object orientation, taking advantage of the more intense downward flux of muons, producing planar imaging with some depth-of-field information in the third dimension. Several simulation studies have been published with multi-orientation tomography, which can form a three-dimensional representation faster than a single orientation view. In this work we present the first experimental multiple orientation muon tomography study. Experimental muon-scatter based tomography was performed using a concrete filled steel drum with several different metal wedges inside, between detector planes. Data was collected from different detector-object orientations by rotating the steel drum. The data collected from each orientation were then combined using two different tomographic methods.

Results showed that using a combination of multiple depth-of-field reconstructions, rather than a traditional inverse Radon transform approach used for CT, resulted in more useful images for sparser data. As cosmic ray muon flux imaging is rate limited, the imaging techniques were compared for sparse data. Using the combined depth-of-field reconstruction technique, fewer detector-object orientations were needed to reconstruct images that could be used to differentiate the metal wedge compositions.

## I. INTRODUCTION

Cosmic ray muons provide a very highly penetrating source of radiation that can be used for imaging large and dense objects. As charged particles ionization detectors can be used to determine the position and direction of the muons. Small angle scattering, and occasionally attenuation, measurements give information on the materials and thicknesses that the muons interact with. Muons have been used for overburden imaging from boreholes [1], [2] and of course pyramid imaging [3], [4] and even volcano imaging [5], [6], by comparing flux rates at different angles due to attenuation.

With detectors placed on either side of an object, particle direction changes can be used to extract scattering information within an object. This can be used to produce a scattering intensity image within the object, where intensity differences are caused by density and atomic number differences. Studies were published by Los Alamos National Laboratory's (LANL) Threat Reduction Team over the evaluation of cosmic-ray muons for this scatter imaging, enabling more sensitive imaging [7], [8], [9], beginning a flurry of activity in applications of the technology. Such imaging is typically performed with a single detector-object orientation taking advantage of the higher downward flux of cosmic ray muons which are created in the upper atmosphere. These have been used in three-dimensional imaging with some depth-of-field information in the vertical direction using the small range of incident zenith angles, producing single orientation tomography. Muon based single orientation tomography has been applied to cargo inspection [10], [11], [12] and high atomic number material detection [13], and even reactor imaging [14], [15].

Proof of concept experiments on large objects were performed. The LANL Threat Reduction Team applied the scatter tomography technique to image inside a Westinghouse MC-10 spent nuclear fuel storage cask [16], [17]. With the large cask size, 2.7 m outer diameter and 4.8 m height [16], [18], the detector panels were on opposite sides of the cask with a large spacing between them. Due to physical constraints, only a small vertical offset of 1.2 m between the detectors at 2.7 m distance was used. This allowed imaging from several different detector-object orientations with the same incident muon flux, multi-orientation tomography, but the large zenith angle limited the observed muon flux to nearly horizontal muons that passed through both detectors, resulting in very low statistics [17] and only a useful a one-dimensional reconstruction. It is important to be able to perform this multi-dimensional imaging with sparse data.

Several simulations were performed by a number of groups on muon tomography of objects from multiple detector-object orientations, up to a full 360° around the object, including used fuel storage casks [19], [20], [21]. These produced very clear tomographic images, but they all used many different detector-object orientations with very high muon statistics.

Our goal in this experimental work was to use multiple detector-object orientation muon scatter data sets to create tomographic images, and then examine the image quality for different levels of statistics – from number of orientations to number of muons per orientation – to enable faster useable field measurements. Data from experimental laboratory measurements is much more accessible than cask measurements, so laboratory sited measurements were performed by the LANL authors. A steel drum was filled with concrete, with metal wedges of different sizes and compositions placed inside, described in the experiment section. The steel drum was placed between the muon detector planes and imaged with different detector-object orientations by rotating the steel drum for different data sets. The image reconstruction of these metal wedges forms the basis of the analysis performed by the UNM authors. These data sets were used to compare the performance of backprojection tomography and a new technique, combined depth-of-field imaging, with varying degrees of data quality by reducing orientations and numbers of muons used

## II. MUON INTERACTIONS

Muons are created in the upper atmosphere by high energy charged particles interacting with air. The muon shower intensity approximately follows a cosine squared function with the zenith angle [22], The muons at larger zenith angles also travel through greater lengths of air from the interaction point to the detector, preferentially attenuating lower energy muons [23].

When passing through matter, muons experience multiple Coulomb interactions resulting in many scatters, and it is the total scattering angle between detectors that we use for analysis. For a single energy muon, total scatter closely follows a Gaussian distribution with a mean of zero radians. The width of the scatter distribution in radians is given by [7]

$$\sigma = \frac{13.6\ MeV}{\beta c p} z \sqrt{\frac{L}{L_0}} [1 + 0.038 \ln(\frac{Lz^2}{L_0 \beta^2})]. \quad (1)$$

This width decreases at higher energies following the muon velocity ($\beta c$) and momentum (p). The radiation length, $L_0$, is material dependent, and the width increases rapidly with an increase in the material atomic number, Z [7], and density. The muons interact with all materials along their paths, so the measured scatter is an integral of the scattering along the entire path. In our data, the muons that pass through a metal wedge also pass through the surrounding concrete, the scatters cannot be individually isolated.

For distributions of energies, the scatter distribution width is a sum of the distributions from different energies and the trend remains: higher scattering angles for higher atomic number and denser materials. Since the test object was rotated in relation to the imaging system, the muon spectrum remained constant throughout the entire measurement campaign, and the specific energies did not have to be accounted for when comparing scattering.

## III. MULTI-VIEW MUON TOMOGRAPHY STUDY

### A. Experiment

Muon track data was collected by the Los Alamos National Laboratory authors in 2017 while analysis was performed by the UNM authors beginning in 2022. Some of the experimental details were not preserved, and so wedge sizes were determined by comparing experimental and simulated imaging, iterating the simulations. The delay between data collection and analysis also prevented feedback into the experiment for image improvement, such as detector gas tube response recalibration.

The experiment measured muon deflection through a single-walled steel drum filled with concrete using Quikrete concrete mix, with lead, tungsten, and brass wedges placed at different positions within the steel drum. Construction of the steel drum and subsequent measurements were performed by LANL authors. The steel drum had a 0.9 mm thick steel wall and the concrete filled volume was approximately 42 cm long and 34 cm in diameter, consistent with a 10-gallon steel drum. The metal wedges were triangular, each with uniform thickness. The brass wedge had a 9 cm base, a height of 7 cm, and a thickness of 5 cm. The lead wedge had a 6 cm base, 7 cm length, and 3 cm thickness. The tungsten wedge had a 15 cm base, a height of 10 cm, and thickness of 4 cm. The lead wedge was centered 20 cm from one end, while the brass and tungsten wedges were coplanar and centered at 32 cm from that same end, or 10 cm from the opposite end. Defining the x direction along the barrel symmetry axis, all wedges had their short edges in the x direction, for wider tomographic views. As muon scatter is atomic number (Z) and density dependent, this range of high Z wedges with varying densities was chosen to test the ability to differentiate these objects from the low Z and low-density concrete surrounding the wedges. Different material densities also affect scatter lengths, so both the Z and density are noted for tungsten (Z=74, $\rho$=19.3 g/cm$^3$), lead (Z=82, $\rho$=11.3 g/cm$^3$), brass (Z=29 and 30, $\rho$=8.5 g/cm$^3$), Quikrete concrete (mix, with principal components Z=8 and 14, $\rho$=2.3 g/cm$^3$), and air (Z=7 and 8, $\rho$=0.96 mg/cm$^3$ at LANL).

The detector array used in this experiment, the Mini Muon Tracker (MMT), Fig. 1, is described in [16], [17], [24], [25]. The MMT was designed by Los Alamos National Laboratory and constructed in collaboration with Decision Sciences International Corporation.

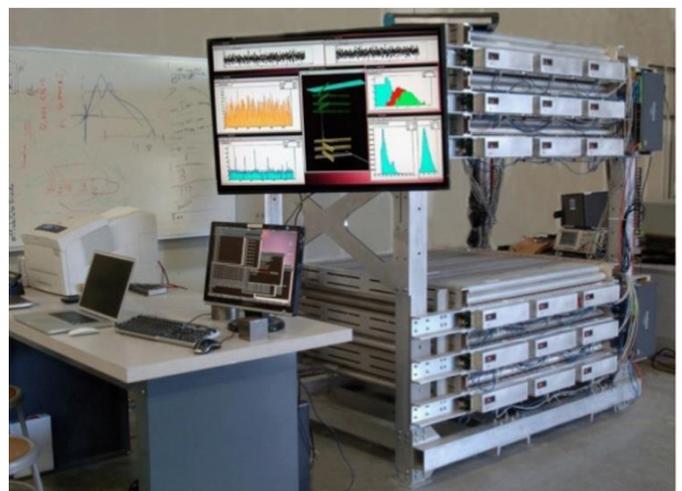

FIG. 1. MMT detectors used for measurements, from [26].

The system consists of two modules each with 12 layers of sealed gas ionization tubes. Tubes contain 47.5% Ar, 42.5% CF$_4$, 7.5% C$_2$H$_6$, and 2.5% He, all % by weight, at 1 atm. They measure 5 cm in diameter with a central anode wire held at

+2550 V, relative to the grounded Al tube wall. Each of the planes consists of parallel tubes, with the layers stacked in perpendicular orientations to gather both x and y information on muon tracks. Interactions in several detector layers were used to determine the trajectory of each muon. The modules were each 1.2 m x 1.2 m in area by 60 cm thick. Two detector modules were used to total 24 layers, with one placed above the other to make use of the high near vertical muon fluence rates. A 60 cm spacing between the detector near faces was used, with the drum (not shown in Fig. 1) placed between the detectors. To avoid edge scatter in/out effects, the detectors were configured to have an active imaging volume of 100 x 100 x 60 cm$^3$ [26].

The read-out electronics, designed by Decision Sciences, utilized tube pulses to determine muon position and reconstruct trajectories through each of the modules. Pulse shapes were used in the reconstruction software to determine muon track position on the millimeter level. For muon signals firing in both modules within the 100 ns coincidence window, the muon position and angle in the top and bottom modules were recorded, along with a time stamp. The on-board FPGAs, with Decision Sciences proprietary programming, handled all pulse processing, and only the positions and angles in each array could be accessed by the user. The seven-column data set was comprised of a time stamp; x, y, and z positions for each module; and the cosine of the angles from each axis for each module. The x, y axes were in the horizontal plane, with x along the drum symmetry axis, and the z axis was vertical.

The scattering angle, $\theta_s$, extracted in this work was calculated using the change in the incoming and outgoing measured muon trajectories, $v_1$ and $v_2$, by

$$cos\theta_s = \frac{\vec{v_1} \cdot \vec{v_2}}{|\vec{v_1}| \cdot |\vec{v_2}|}. \qquad (2)$$

The components of each vector along the x y and z axes are given by the cosine values in the data set.

Each scatter angle was used along each corresponding muon track to populate the image space using two different imaging methods. Since scatters are typically small (muons through the strongest scatterer, tungsten, experienced a 71.6 mrad (4.1°) average scatter) and we wanted to compare the imaging methods directly without complications from spline fitting or point of closest approach scatter position determination from incoming and outgoing vectors, a simple straight-line projection from the top detector was used for the muon track through the imaging space.

The steel drum was held in an aluminum cradle that allowed rotation and manually rotated in 15-degree ($\pi$/12 rad) steps to better than 1 degree and a new data set was recorded for each orientation. This process was repeated for a full 360-degree rotation, resulting in 24 total measurement data sets around the drum. The chosen 15-degree steps provide an abundance of detector-object views that allowed for incremental reductions in the amount of data utilized for reconstructions. We refer to drum-detector orientations in degrees for conceptual clarity, but muon scattering calculations were performed in radians.

The average recorded muon event rate was 33 muons/sec, and most orientations were measured for just under 1 day. This is lower than the rate of muons passing through the active region since the array was set to veto background gammas and neutrons for prior spent fuel cask measurements by requiring neighboring tubes to fire in coincidence. This is in addition to requiring tracks to be established in both the upper and lower module within the 100 ns coincidence window by the tubes that did fire. The minimum number of recorded muon tracks was 2.4 million for one of the shorter runs over 20 hours, and the maximum was 5 million tracks over 2 days at the 0-degree orientation to allow for a better single orientation analysis. For consistency, 2.4 million muon tracks were used from each orientation when performing multiple orientation imaging.

An important feature of this approach to imaging is that, since the steel drum was rotated but the detector did not move, every detector-object orientation has the same muon fluence rate and energy spectrum. This is actually similar to fuel cask data using detectors to the sides of the cask [16], [24], where the different views are from different azimuthal angles though the zenith angle of acceptance did not change. It is noted that the muon energy spectrum and intensity change with zenith angle, so this consistency in measurements allows different orientations to be compared directly, an orientation agnostic approach that allows a traditional tomographic approach to be studied, as in prior simulation studies [19], [21]. Simulations in the current work used a realistic energy spectrum and the angular dependence for direct comparison.

Beyond that, in muon tomographic imaging using a single orientation, a depth-of-field image may be constructed with the three-dimensional information from a set of muon trajectories. The multiple orientation approach with no preferred orientation allows a new method of combining depth-of-field information with equal importance from the different orientations, explained below. Both traditional multiple orientation backprojection tomography and the new combined depth-of-field imaging were examined.

### B. Simulations

Muon scattering simulations were performed using Geant4 [27], [28] version 10.7, using the CRY (cosmic ray yield data) [29] incident cosmic ray muon energy distribution and angular dependence with the location set for Los Alamos, New Mexico. The detectors were modeled as 1.2 m x 1.2 m planes with 60 cm separation. The muon source was distributed uniformly over the top plane. The drum, concrete, air, and wedges were modeled as described in the experimental section, using Geant4 material definitions. The simulated detectors had perfect efficiency and angular resolution.

For muons passing through both detector planes, the positions and trajectories through each detector were recorded to match the 6-column position and angle data format of the Decision Sciences software output. Image reconstruction analysis was performed exactly as the experimental data.

To determine wedge sizes, simulations were repeated, and the wedge sizes were modified until image reconstructions from simulation matched experimental reconstructions. For comparison with experiment, 24 orientations were used at 15-degree steps, with the first 2.4 million tracked muons (tracked in both top and bottom detectors) used in image reconstructions.

## IV. MULTI-ORIENTATION ANALYSIS METHODS

## A. Sinogram and inverse Radon transform reconstruction

Following methods from medical imaging, the first technique utilized for tomographic reconstruction involved the population of a sinogram and subsequent backprojection via inverse Radon transform [30], [31]. This follows the backprojection method applied in previous muon tomography simulations [20].

The centerline of the drum was used for the image reconstruction rotational axis. Using the muon trajectory and position recorded from the top detector, the distance of closest approach to the centerline, r, was found and the angle between the projected trajectory and the normal line that r lies on, θ, were extracted, Fig. 2. Both r and θ were used to determine the sinogram bin in which the muon scattering angle was tallied. After all muons were projected and tallied the average muon scattering angle in each r, θ position of the sinogram was found. A sinogram was produced for each 1 cm step in x, along the drum symmetry axis. As mentioned, for more direct comparison between image reconstruction techniques, a straight-line trajectory from the top detector was used.

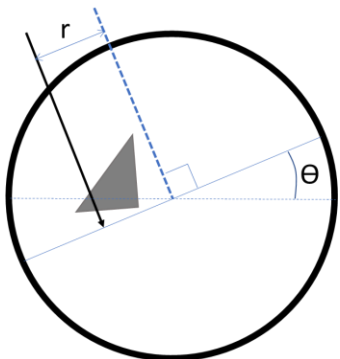

FIG. 2. Sinogram population diagram demonstrating distance of closest approach of muon to drum centerline, r, and angle of the line normal to the muon vector, θ, with a lead wedge and the drum boundary.

Populated sinograms were backprojected using the iradon function in MATLAB [32] to construct the image, populating 1 cm$^3$ voxels in the imaging space. While the sinogram intensities in each (r,θ) bin were average scattering angles, the resulting backprojected intensity values in each voxel were not directly comparable to the intensities in the combined depth-of-field reconstruction and were normalized for numerical comparisons between methods.

## B. Combined depth-of-field reconstruction

In x ray-based tomography based on photon attenuation, the initial x rays travel in a straight line and the spatial intensity on the detector may be traced back through the object to determine the line along which attenuation occurred, with several orientations used to find the position of attenuation within the object. In muon scatter imaging, even from a single detector-object orientation, the trajectory of the muon is determined, which is a major difference that aids image reconstruction.

As muon trajectories are followed in the image space, the paths with high scatter converge, which for muon imaging corresponds with the position of denser materials and/or higher atomic numbers. By tallying the scattering angle of each muon along its path through a voxelated imaging volume, a three-dimensional image of the scattering centers was produced. The resulting image, from a single detector-object orientation, thus also provides depth information. A diagram of this tally structure is shown in Fig. 3. Black tracks represent muons that slightly scatter via interactions only with air. Orange tracks represent muons that only interact with the steel drum and concrete within, resulting in larger scattering angles than muons that only interact with air. Red tracks represent muons that experience relatively large scattering events during interactions with the high Z wedges, and these tracks converge on the wedge. Due to the small angular spread in muon trajectories, there is depth information (z), along with the lateral information (x, y) of projection imaging. To keep data consistent between the imaging methods, a straight trajectory from the top detector was used for reconstruction.

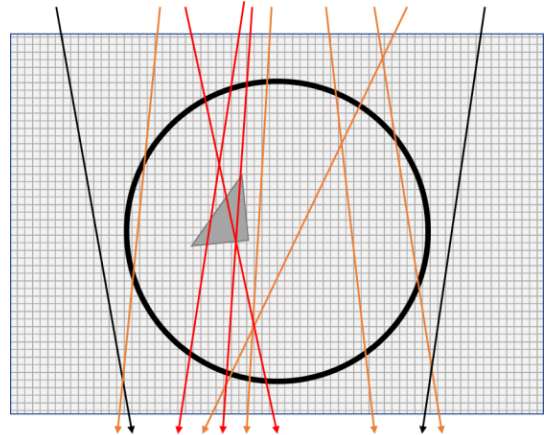

FIG. 3. Projected tracks from the top detector of muons, with red having a high scattering angle, orange medium, and black low. The high scatter muon paths converge on the metal wedge, producing depth information from the muon paths. The scattering angle of each muon that passes through any voxel was recorded and a path length weighted average was tallied for each voxel. We used straight tracks from the top detector to keep data consistent between imaging techniques.

In our implementation of depth-of-field image reconstruction, the three-dimensional imaging volume consisted of 1 cm$^3$ voxels. The scattering angle for each muon passing through a voxel was tallied for that voxel and weighted by the path length through the voxel. As scattering is an integral of muon interactions in the material over the full path of travel, muons with a longer path within a voxel will sample the material in the voxel more than muons that pass through a smaller section of the voxel. This allowed for the calculation of a path length weighted average scattering angle in each voxel.

The range of muon angles was limited to those that passed through both detectors, which reduced image resolution along the depth (z) direction between detectors when using a single orientation. As the steel drum was rotated, the range of observed muon angles relative to the steel drum increased and improved position resolution along what was previously the z axis of the barrel. As the orientation is relative, in analysis the steel drum position was fixed in the imaging voxel space, and the trajectories of the muons were rotated accordingly. This allowed for a single voxelated tally volume to be populated where the muon vectors were mapped onto the fixed imaging space voxel grid.

The depth-of-field images acquired at several different detector-object orientations were combined [33] and expected to converge to a three-dimensional image more quickly than traditional tomography, which is important with sparse data scenarios.

The essential differences between imaging using the backprojection and combined depth-of-field imaging techniques may be summarized. For backprojection, prior to the inverse Radon transform the average scattering angle was found for each projection axis position r and axis angle θ, and then projected to map onto the volume voxels. for depth-of-field imaging this was averaged per voxel in the imaging volume directly. In reconstructions, these differences manifested as differences in the observed image intensities and expected image streaking for backprojection with sparse data.

## V. MULTIPLE ORIENTATION IMAGING RESULTS AND DISCUSSION

### A. Images

Three-dimensional images based on the combined depth-of-field technique using all 24 available detector-object orientations are shown in Fig. 4. In Fig. 4(a), image intensity directly relates to the average scattering angle within each voxel. A minimum average scatter intensity threshold of 42 milliradians was selected by visual inspection to omit voxels outside of the steel drum that consist of air. In Fig. 4(b), a minimum threshold of 47.5 milliradians was selected by visual inspection to highlight the metal wedges. A slight deformation is seen in the side of the steel drum image due to muon shadowing effects from the high Z, high density tungsten wedge. This same shadowing effect is the cause of the high intensity points seen between the wedges in Fig 4(b). Expected average scattering values for different materials are discussed below.

Figure 5 shows a coronal slice, perpendicular to the steel drum symmetry axis along a single value of x, through the lead wedge. A similar coronal slice through the tungsten and brass wedges, which are coplanar, is shown in Fig. 6. This two-dimensional slice was used for the tungsten and brass imaging analysis in this work. Analysis of the lead slice is omitted for brevity but is discussed in [34]. As mentioned, the muon scatters along the entire path, so concrete scattering includes effects from air and the thin barrel wall, and the wedges include these along with the wedge materials.

Tomographic images that correspond to the coronal slice in Fig. 6 were created for the image quality comparisons. As a best-case scenario, 2.4 million muons were utilized for each detector-object view, with 15-degree steps through 360 degrees for 24 views, Fig. 7. In both images the round concrete can be seen, the tungsten wedge is very clear in the bottom left, and the brass wedge can also be seen in the upper right. The outer bounds of the concrete can be discerned more clearly from the surrounding air for the backprojection image. However, the tungsten and brass wedges are more greatly contrasted from the surrounding concrete in the combined depth-of-field reconstruction. For examining metals, the combined depth-of-field results are more favorable. Image quality metrics were found and are presented following the comparison images.

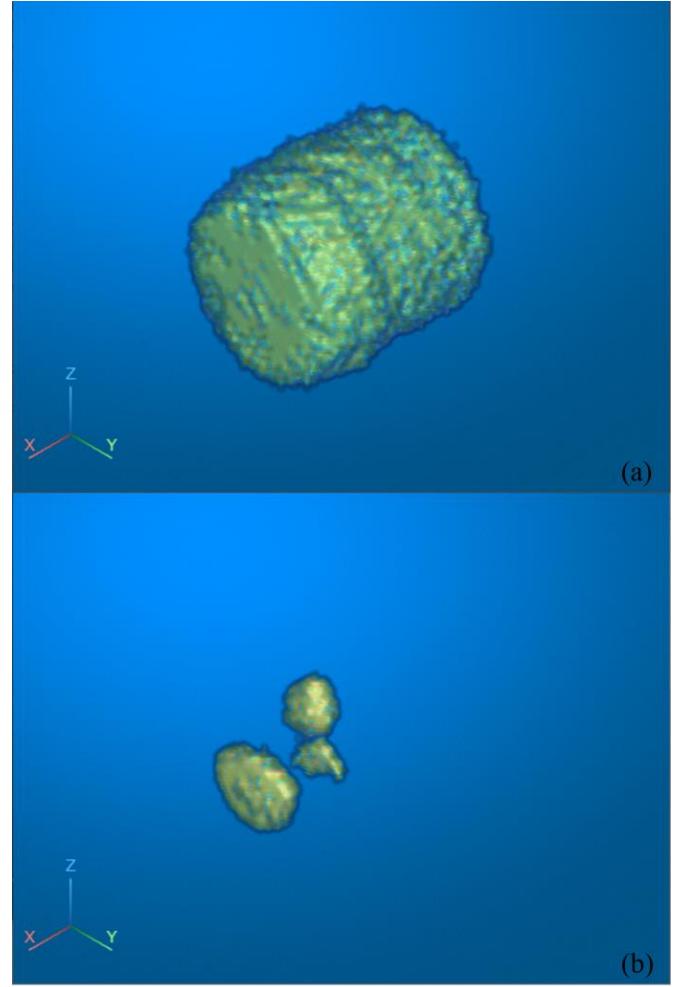

FIG. 4. (a) 3D reconstruction of steel drum exterior with a voxel threshold at 42 mrad and (b) of metal wedges with a voxel threshold at 47.5 mrad. The bottom left wedge is tungsten and the bottom center wedge is brass, and on the top in a separate x plane is lead.

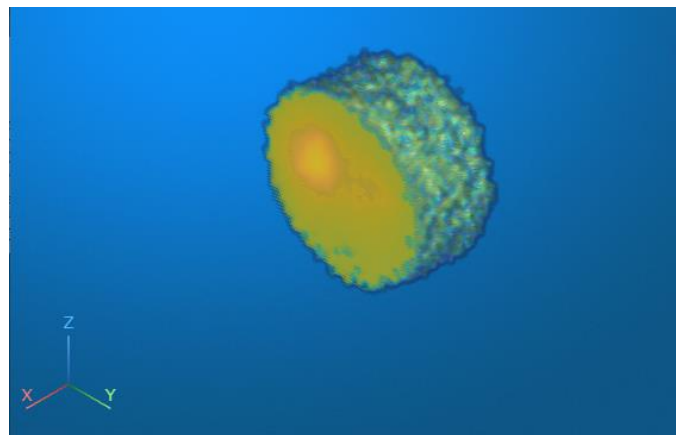

FIG. 5. 3D reconstruction of steel drum with coronal slice along a single x through lead wedge.

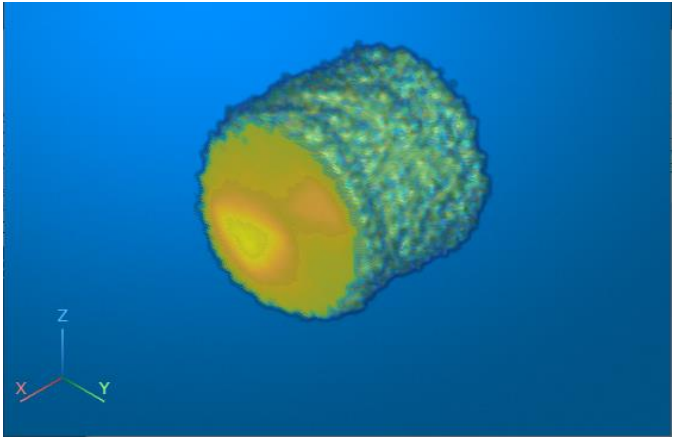

FIG. 6. 3D reconstruction of steel drum with coronal slice along a single x through tungsten and brass wedges.

orientations and only 1 million muons per orientation, a total of 2 million muons vs. 57.6 million for Fig. 7. In the combined depth-of-field image, the limits of the tungsten wedge towards the bottom left are visible, and the brass wedge still appears. For the backprojection technique, the streaking broadens the wedges, making localization more difficult, and the poor contrast from the surrounding concrete makes the materials more difficult to distinguish.

In order to evaluate the quality of reconstructed images and compare the performance between similar techniques quantitatively, techniques common to CT image analysis were applied: quantifying the contrast resolution and the spatial resolution. For contrast resolution, the contrast-to-noise ratio was used, and for spatial resolution the image sharpness was examined.

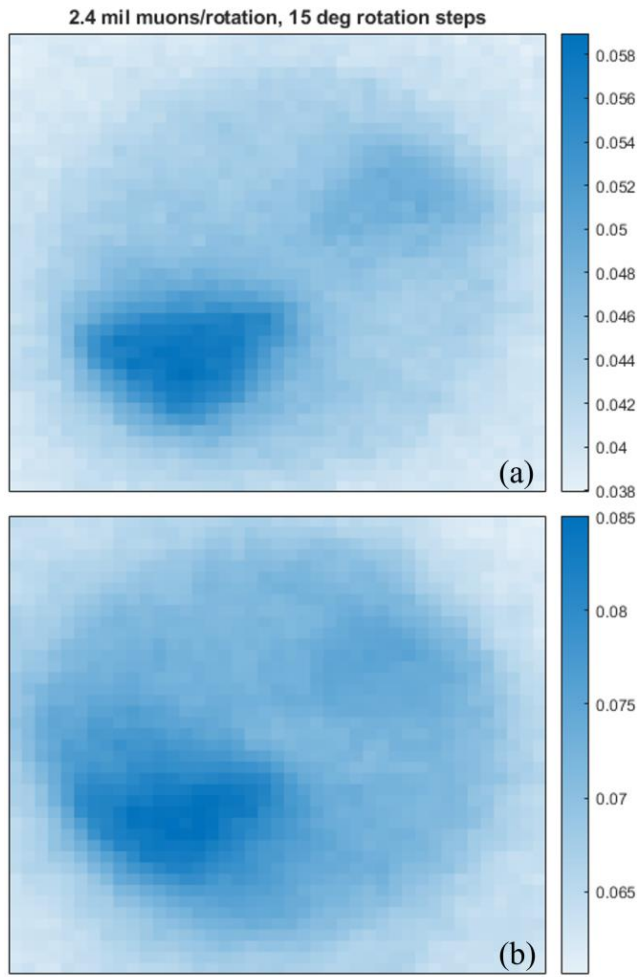

FIG. 7. Image reconstructions using combined depth-of-field (a) and backprojection (b) using 24 orientation views and 2.4 million muons per orientation.

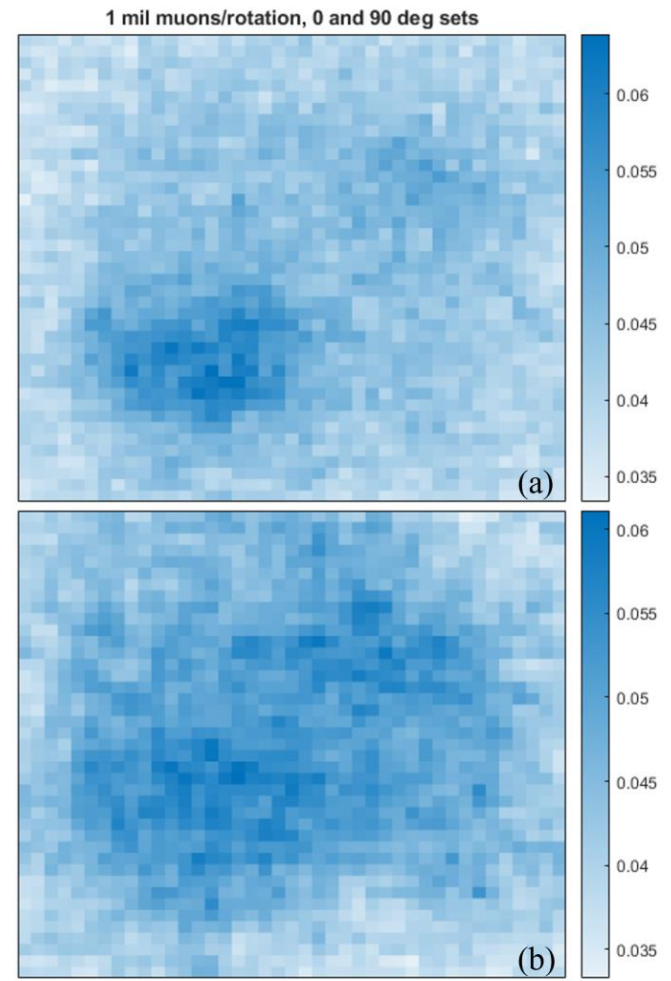

FIG. 8. Image reconstructions using combined depth-of-field (a) and backprojection (b) using 2 different detector-object orientations views and 1 million muons per orientation.

To study the reconstruction performance in less ideal scenarios with sparse data, additional reconstructions were performed on a small subset of the data.

The reconstructions in Fig. 8 were performed with two different steel drum rotation data sets at 0- and 90-degree

### B. Contrast resolution

For a particular reconstruction image, the observed intensity distribution of each material was extracted by isolating the relevant region in the image. Using these distributions, the average material intensity value (μ) and the standard deviation

(σ) were extracted for each material in an image. The contrast-to-noise ratio (CNR) can be calculated using,

$$CNR = \frac{(\mu_1 - \mu_2)}{\sigma_2}. \qquad (3)$$

following the American College of Radiology QC guidelines for CT [35]. Here $\mu_1$ is the average intensity of the foreground material, for example a wedge, and $\mu_2$ and $\sigma_2$ are the average intensity value and standard deviation of the background material, for example the surrounding concrete. For concrete CNR evaluations, concrete was the foreground material and air was the background material. The μ and σ for reconstructions shown in Fig. 7 are presented below in Table I. The values shown in Table II were extracted from Geant4 simulations of an identical slice. The values in Table I and II were then scaled and normalized from 0 to 1 to create the plot shown in Fig. 9.

TABLE I. Average intensity and standard deviation values from experimental reconstructions created with all 24 unique detector-object orientations, 2.4 mil muons per view, as in Fig. 7.

|  | Backprojection | Depth-of-field |
|---|---|---|
|  | Intensity (arb) | Intensity (rad) |
| Tungsten | 0.0815 ± 0.0026 | 0.0548 ± 0.0025 |
| Brass | 0.0743 ± 0.0008 | 0.0482 ± 0.0006 |
| Concrete | 0.0723 ± 0.0028 | 0.0445 ± 0.0024 |
| Air | 0.0618 ± 0.0015 | 0.0383 ± 0.0007 |

TABLE II. Average intensity and standard deviation values from simulated reconstructions created with all 24 unique detector-object orientations, 2.4 mil muons per view, equivalent to Fig. 7.

|  | Backprojection | Depth-of-field |
|---|---|---|
|  | Intensity (arb) | Intensity (rad) |
| Tungsten | 0.0641 ± 0.0046 | 0.0446 ± 0.0033 |
| Brass | 0.0453 ± 0.0009 | 0.0306 ± 0.0007 |
| Concrete | 0.0430 ± 0.0057 | 0.0250 ± 0.0045 |
| Air | 0.0209 ± 0.0025 | 0.0110 ± 0.0013 |

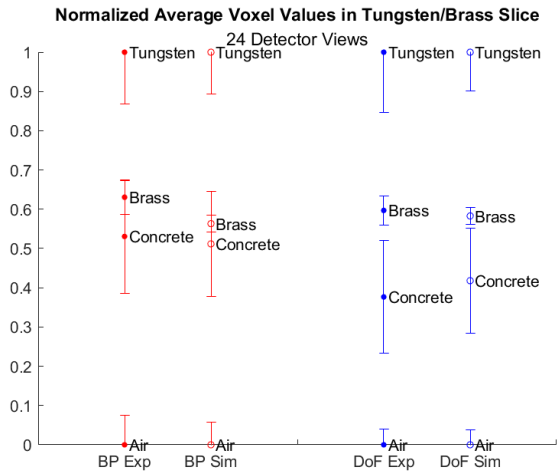

FIG. 9. Average intensity values, with error bars representing the standard deviation of intensity values for a particular material. Solid points represent experimental values and empty points represent values extracted from idealized simulations.

Using the values in Table I and Table II, CNR values were calculated using Equation 3 and the resulting values are shown in Table III. Larger CNR values are desirable and quantify a greater distinction of a particular object/material from its surroundings. In both experimental and simulated reconstructions, the combined depth-of-field technique shows greater contrast for each of the three wedges from the concrete.

TABLE III. Experimental and Simulated CNR values from reconstructions created with all 24 unique detector-object orientations, 2.4 mil muons per view, calculated using values in Table I and Table II.

| Foreground-Background | Exp Back-Projection | Sim Back-Projection | Exp Depth-of-field | Sim Depth-of-field |
|---|---|---|---|---|
| Tungsten-Concrete | 3.25 | 3.67 | 4.35 | 4.37 |
| Brass-Concrete | 0.69 | 0.39 | 1.54 | 1.24 |
| Concrete-Air | 7.15 | 8.89 | 9.36 | 11.20 |

The same analysis was repeated for the sparse data reconstructions shown in Fig. 8 based on two view orientations and 1 million muons per view. The results are shown below in Table IV and were normalized and scaled to 0 to 1 to create Fig. 10. CNR values from this two-view reconstruction are shown in Table V. Visually, tungsten is still very clearly distinguishable from surroundings using the depth-of-field technique, and the difference between tungsten and concrete is larger than in the backprojection technique. The CNR values quantify this, with a tungsten-to-concrete CNR of 2.91 for combined depth-of-field compared with 1.27 for backprojection. This is important to note for applications where the goal is to identify and distinguish materials with high density and high atomic numbers from lower density and lower atomic number materials. Simulated results show a larger performance gap between the two techniques. While the experimental performance was good, the detector gas has degraded following previous measurements on the spent fuel storage cask, and the pulse shape response fit to extract mm scale resolution may have suffered. As angular resolution in the system is restored, such as with a change out of gas, we expect performance to improve toward the simulated results.

TABLE IV. Average intensity and standard deviation values from experimental reconstructions created with 2 unique detector-object orientations, 1 mil muons per view, as in Fig. 8.

|  | Backprojection | Depth-of-field |
|---|---|---|
|  | Intensity (arb) | Intensity (rad) |
| Tungsten | 0.0537 ± 0.0031 | 0.0547 ± 0.0039 |
| Brass | 0.0532 ± 0.0027 | 0.0481 ± 0.0027 |
| Concrete | 0.0483 ± 0.0043 | 0.0445 ± 0.0035 |
| Air | 0.0357 ± 0.0036 | 0.0373 ± 0.0025 |

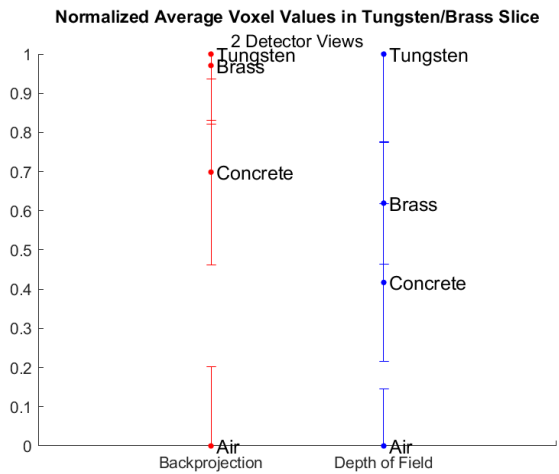

FIG. 10. Average intensity values, with error bars representing the standard deviation of intensity values for a particular material. Solid points represent experimental values and empty points represent values extracted from idealized simulations.

TABLE V. Experimental and Simulated CNR values from reconstructions created with 2 unique detector-object orientations, calculated using values in Table IV.

| Foreground-Background | Exp Back-Projection | Sim Back-Projection | Exp Depth-of-field | Sim Depth-of-field |
|---|---|---|---|---|
| Tungsten-Concrete | 1.27 | 2.84 | 2.91 | 4.27 |
| Brass-Concrete | 1.14 | 0.92 | 1.01 | 1.20 |
| Concrete-Air | 3.47 | 6.13 | 2.87 | 7.04 |

### C. Spatial resolution

In computed tomography applications, line pair phantoms/gratings with different spacings are imaged to find the spatial resolution of the system [36]. Without a grating, as in this study, the edge rise distance can be used. This distance is found between a neighboring low and high intensity region, examining the pixel distance between 10% and 90% intensity levels in the image (e.g., [37-39]). Since the muon scatter is due to all materials along its path, voxel intensities near high Z and high-density materials may be higher due to a shadowing effect onto those voxels. In order to compare spatial resolution, one-dimensional intensity slices were taken from the reconstructed images and image sharpness compared. An example slice is taken through tungsten, but also containing concrete and air, from the 24-orientation data used in Fig. 7. Fig. 11 shows the slice line in images from both imaging modalities.

The intensity profiles are presented in Fig. 12, where vertical position refers to the pixel position in the z direction from the center of the tungsten, along the line marked in Fig. 11. The average tungsten and concrete values from Table I are chosen to be 1 and 0, respectively, for easier comparison between images. As seen in Fig. 12, there is little difference in the spatial resolution between the two reconstruction techniques. The edge rise distance for both techniques being approximately 7 pixels for the left boundary between concrete and the tungsten wedge in both techniques. On the right-side boundary between the wedge and concrete, the edge rise distance is approximately 3 pixels for both techniques.

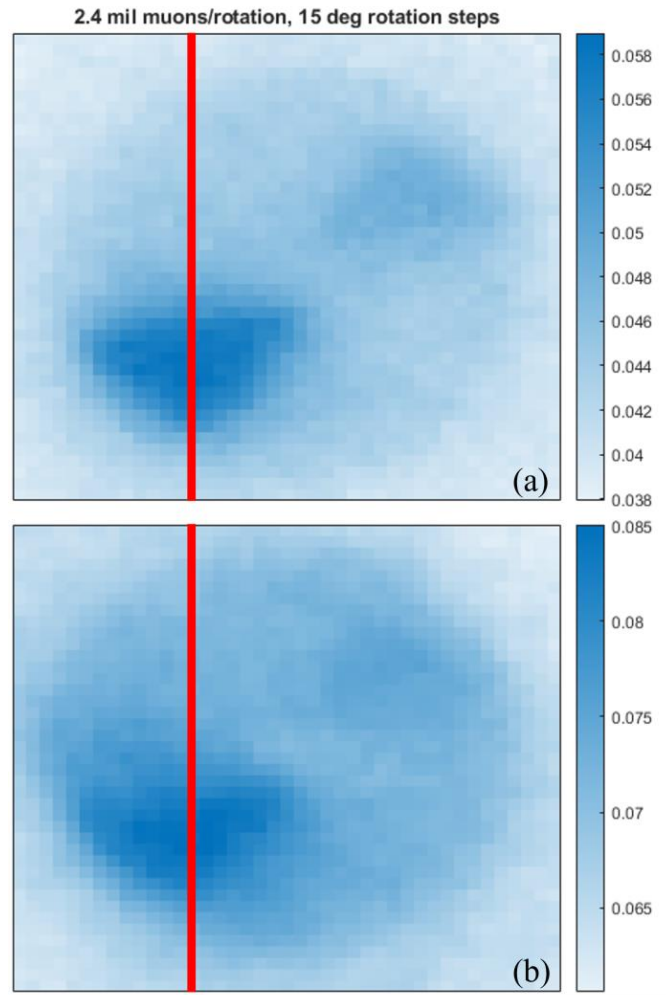

FIG. 11. Line shows slice region used to create contrast plots. An identical slice is selected from each analysis method. Combined Depth-of-field (a), backprojection (b), from Fig. 7.

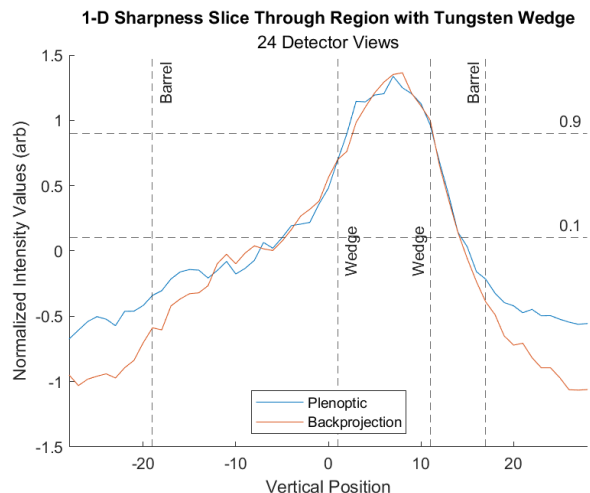

FIG. 12. Intensity profiles for combined depth-of-field and for backprojection imaging through line in Fig. 11.

## V. CONCLUSION

Multiple orientation experimental muon tomography studies were performed using a concrete filled steel drum with tungsten, lead, and brass wedges. Data were taken at 24 different orientations in 15-degree steps, with 2.4 million muon tracks from each orientation. The data was agnostic as to view orientation, there is no difference in incident muon flux for the different views. In this work, the steel drum was rotated while the detectors faces were kept horizontal, and the near-vertical cosmic ray muon flux was used. This can be applied to other view agnostic imaging such as spent fuel storage casks in which views are taken from different azimuthal angles and a constant zenith angle, as was demonstrated in prior simulation studies. This approach allowed us to use tomography techniques that use many different angles, both backprojection using inverse Radon transform and a new combined depth-of-field approach were investigated, to produce three dimensional images. While simulated reconstructions had been performed before on multiple orientation muon tomography, this was the first experimental study, and the first to apply a combined depth-of-field approach.

The backprojection and combined depth-of-field approaches showed similar spatial resolution performance while there was ample data, 24 orientations and 2.4 million muons per orientation. However, the combined depth-of-field technique showed significantly better contrast resolution. Sparse data using fewer orientations and muons were examined to understand the effect on image quality with more realistic field measurement limitations. The combined depth-of-field technique showed great advantage over the backprojection technique in contrast resolution, more easily distinguishing high-density materials from concrete.

## ACKNOWLEDGEMENTS

This work is supported by the National Nuclear Security Administration's Office of Defense Nuclear Nonproliferation Research and Development. This work is also supported by the Consortium for Monitoring, Technology, and Verification under Department of Energy National Nuclear Security Administration award number DE-NA00003920. We also acknowledge the support of the Los Alamos National Laboratory LDRD program funding for this work.